# Exceedingly large in-plane critical field of finite-momentum pairing state in bulk superlattices

Junfa Lin[1,#], Ziqiao Wang[2,7,#], Hongyi Yan[3,10,#], Xiaoping Ma[1,4,#], Zihan Cui[1], Yu Zhang[1], Yuxuan Lei[2], Jie Liu[1], Rao Li[1], Chuanying Xi[5], Zengwei Zhu[6], Huakun Zuo[6], Yanzhao Liu[7], Huaixin Yang[4], Tian-Long Xia[1,8,9,*], Haiwen Liu[3,10,11,*], Yi Liu[1,8,*], and Jian Wang[2,12,13,*]

[1] *School of Physics and Beijing Key Laboratory of Opto-electronic Functional Materials & Micro-nano Devices, Renmin University of China, Beijing 100872, China.*

[2] *International Center for Quantum Materials, School of Physics, Peking University, Beijing 100871, China.*

[3] *Center for Advanced Quantum Studies, School of Physics and Astronomy, Beijing Normal University, Beijing 100875, China.*

[4] *Beijing National Laboratory for Condensed Matter Physics and Institute of Physics, Chinese Academy of Sciences, Beijing 100190, China.*

[5] *Anhui Key Laboratory of Low-Energy Quantum Materials and Devices, High Magnetic Field Laboratory, HFIPS, Chinese Academy of Sciences, Hefei, Anhui 230031, China.*

[6] *Wuhan National High Magnetic Field Center, Huazhong University of Science and Technology, Wuhan 430074, China.*

[7] *Quantum Science Center of Guangdong-Hong Kong-Macao Greater Bay Area (Guangdong), Shenzhen 518045, China.*

[8] *Key Laboratory of Quantum State Construction and Manipulation (Ministry of Education), Renmin University of China, Beijing 100872, China.*

[9] *Laboratory for Neutron Scattering, Renmin University of China, Beijing 100872, China.*

[10] *Key Laboratory of Multiscale Spin Physics (Ministry of Education), Beijing Normal University, Beijing 100875, China.*

[11] *Interdisciplinary Center for Theoretical Physics and Information Sciences, Fudan University, Shanghai 200433, China.*

[12] *Hefei National Laboratory, Hefei 230088, China.*
[13] *Collaborative Innovation Center of Quantum Matter, Beijing 100871, China.*

#These authors contribute equally: Junfa Lin, Ziqiao Wang, Hongyi Yan, Xiaoping Ma

*Correspondence to: jianwangphysics@pku.edu.cn (J.W.), yiliu@ruc.edu.cn (Y.L.),

haiwen.liu@bnu.edu.cn (H.L.), tlxia@ruc.edu.cn (T.-L.X.)

**Magnetic flux profoundly influences the phase factor of charge particles, leading to exotic quantum phenomena. A recent example is that the orbital effect of magnetic field could induce finite-momentum pairing state in nanoflakes, which offers a new pathway to realize the spatially modulated superconductivity distinct from the Fulde–Ferrell–Larkin–Ovchinnikov (FFLO) state induced by Zeeman effect. However, whether such intriguing state can exist in the bulk materials under extremely large magnetic field remains elusive. Here we report the orbital effect induced finite-momentum pairing state with exceedingly large in-plane critical field in a bulk superconducting superlattice. Remarkably, the in-plane critical field shows a pronounced upturn behavior, exceeding eight times the Pauli limit which is comparable to monolayer Ising superconductor. Under high in-plane magnetic fields, significant anisotropic transport behavior between the interlayer and intralayer directions is detected, highlighting the critical role of suppressed interlayer coherence in the orbital effect induced finite-momentum pairing state. Crucially, this finite-momentum pairing state remains robust against moderate disorder. Our findings suggest that van der Waals superlattices, with strong Ising spin-orbit coupling and tunable interlayer coherence, offer new avenues for constructing and modulating unconventional superconducting states.**

The pursuit for new superconducting materials surviving extremely high magnetic field has been a central issue in both fundamental research and potential applications[1-6]. In general, the superconductivity is destroyed under external magnetic field via the coupling to spin and orbital degree of freedom. In bulk dirty superconductors, the critical field is usually restricted by the Chandrasekhar-Clogston or Pauli paramagnetic limit, $B_P = 1.86T_c$ (T/K) with $T_c$ denoting the superconducting transition temperature[7,8]. Nevertheless, the interaction between large magnetic fields and the spin components of Cooper pairs leads to novel superconducting states that transcend the Pauli limit, manifesting in phenomena such as Ising superconductivity[1-4,9-11], spin-triplet state[5,6,12], and the Fulde–Ferrell–Larkin–Ovchinnikov (FFLO) state[13-17]. The experimental observation of upturn feature in the critical field beyond the Pauli limit at low temperatures provides experimental signatures of the long-sought FFLO state.

Magnetic field introduces a phase factor into the superconducting wave function via the magnetic vector potential, leading to observable quantum effects such as flux quantization and Fraunhofer diffraction patterns in superconducting devices[18]. A recent innovative proposal suggests that the orbital effect of the in-plane magnetic field could induce finite-momentum pairing state in bilayer transition metal dichalcogenides (TMDs)[19,20]. This exotic state involves the modification of magnetic vector potential on the phase factor (i.e. orbital effect), distinct from the FFLO state that stems from the mismatch between the spin-up and spin-down Fermi surface (i.e. Zeeman effect). The Aharonov-Bohm phase $\Delta\varphi = 2e\Phi/\hbar$ under a magnetic flux $\Phi$ (Fig. 1a)[21] results in the momentum shift $\Delta q = 2eBD/\hbar$ between adjacent layers (where $e$, $B$, $D$, $\hbar$ are the electron charge, the external in-plane field, the layer spacing and the reduced Planck constant, see Fig. 1b). Experimental signature of the orbital effect induced finite-momentum pairing state is revealed in bilayer[22] and few-layer TMD materials[23], featured as an upturn behavior in the temperature dependence of the in-plane critical field $B_{c2,\parallel}(T)$. However, whether this intriguing state can exist in the bulk materials under extremely large magnetic field remains an open question.

Here, we report the experimental observation of the orbital effect induced finite-momentum pairing state in a bulk superconducting superlattice $Ba_6Ta_{11}S_{28}$ with strong Ising spin-orbit coupling (SOC) and relatively weak interlayer Josephson coupling. An exceptionally high in-plane critical field exceeding eight times the Pauli limit has been observed, which is among the highest ever reported for bulk superconductors and is comparable to that of monolayer Ising superconductors. The pronounced upturn in $B_{c2,\parallel}(T)$ can be quantitatively explained by the finite-momentum pairing state, which markedly contrasts with uniform superconductivity. Systematic investigations across various disorder strengths demonstrate that the observed finite-momentum pairing states are resilient to moderate disorder scattering. Importantly, targeted measurements highlight significant differences in transport behavior along the interlayer and intralayer directions under high in-plane magnetic fields, underscoring the critical role of suppressed interlayer coherence via the Josephson vortex solid melting in the emergence of this spatially modulated superconducting state.

**Crystal structure and superconductivity of $Ba_6Ta_{11}S_{28}$ superlattice**

We have synthesized a bulk van der Waals superlattice $Ba_6Ta_{11}S_{28}$ crystal using the flux method (see Methods for details). Figure 1c illustrates an atomically resolved high-angle annular dark field scanning transmission electron microscopy (HAADF-STEM) image of the cross section of $Ba_6Ta_{11}S_{28}$ superlattice along the [100] zone axis superimposed by a schematic of lattice structure. One unit cell ($a = b = 9.6$ Å, $c = 24.1$ Å) of $Ba_6Ta_{11}S_{28}$ consists of two inversion-related H-$TaS_2$ layers and the neighboring $Ba_3TaS_5$ block layers. The distance between adjacent H-$TaS_2$ layer is more than doubled that of bulk 2H-$TaS_2$[24], which results in a weaker interlayer coupling in $Ba_6Ta_{11}S_{28}$. Furthermore, a detailed analysis of the HAADF image reveals that the H-$TaS_2$ layers display high quality with a well-ordered lattice structure, while the $Ba_3TaS_5$ layers show disordered Ta atoms (Fig. 1c). The disorder in the $Ba_3TaS_5$ layers is also evidenced by the selected-area electron diffraction (SAED) pattern taken along the [100] zone axis (Supplementary Fig. S1c), where the diffraction points from the block layers blur into elongated streaks. A scanning tunneling microscopy (STM) topography of the H-$TaS_2$ terminated $Ba_6Ta_{11}S_{28}$ is shown in Fig. 1d, further demonstrating the high quality of the H-$TaS_2$ layers. Moreover, previous angle-resolved photoemission spectroscopy (ARPES) measurements[25], in combination with the band structure calculation, indicate that the local density of states (LDOS) of $Ba_6Ta_{11}S_{28}$ around the Fermi level are contributed by H-$TaS_2$ layers whereas $Ba_3TaS_5$ block layers are not involved.

In Fig. 1e, the temperature dependence of resistivity $\rho_{ab}(T)$ for $Ba_6Ta_{11}S_{28}$ shows a metallic behavior with residual resistivity ratio (RRR, defined as $\rho_{ab}(300\,K)/\rho_{ab}(3\,K)$) of 6.4. An enlarged view in the low temperature regime exhibits a sharp superconducting transition (the inset of Fig. 1e), where the resistivity starts to decrease at 2.72 K ($T_c^{\text{onset}}$), and drops to zero within the measurement resolution at 2.62 K ($T_c^{\text{zero}}$). Note that the superconducting critical temperature of $Ba_6Ta_{11}S_{28}$ is comparable to the value of monolayer $TaS_2$ ($T_c^{\text{zero}} = 2.0$ K)[4], and three times higher than that of bulk 2H-$TaS_2$ ($T_c^{\text{zero}} = 0.8$ K)[26]. Moreover, the zero-field cooling (ZFC) and field cooling (FC) magnetic susceptibility $4\pi\chi$ are measured under in-plane magnetic field. ZFC measurement yields the shielding fraction of 94% at 1.8 K, indicating bulk superconductivity (Supplementary Fig. S2).

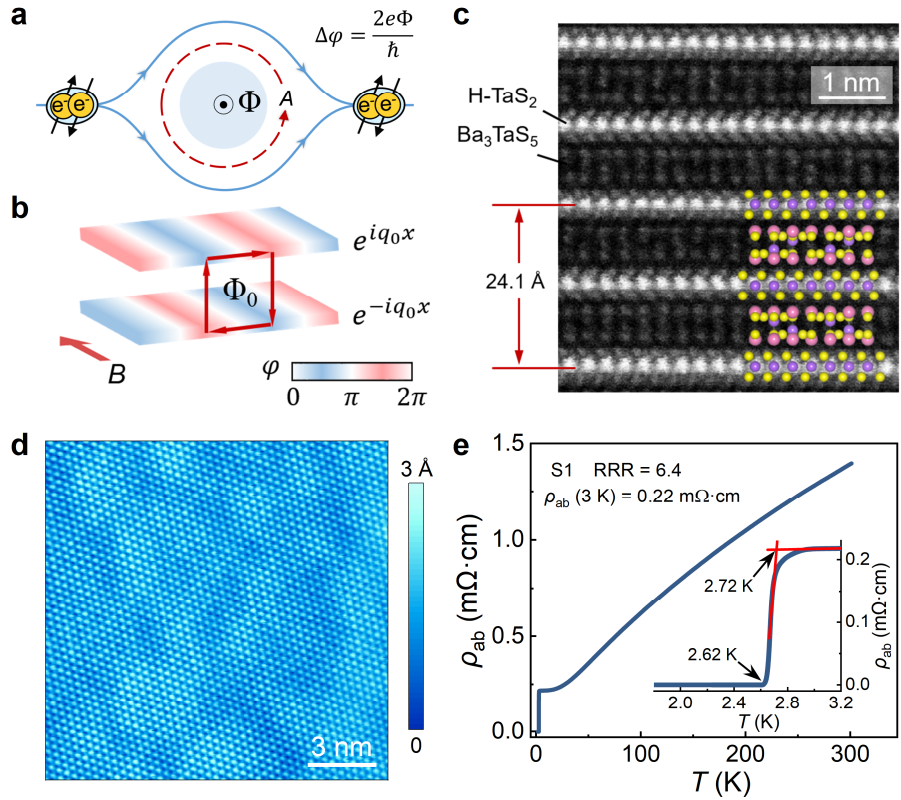

**Fig. 1 | Crystal structure and superconductivity of $Ba_6Ta_{11}S_{28}$ superlattice. a**, The schematic of the Cooper pair interference under the influence of magnetic flux. When encircling a magnetic flux $\Phi$, the Cooper pair would acquire a phase factor $\exp(i\Delta\varphi) = \exp(i2e\Phi/\hbar)$. **b**, The schematic of orbital effect induced finite-momentum pairing state in a bilayer TMD, a recent example revealing the effect of magnetic field on the phase factor. The orbital effect of in-plane magnetic field leads to spatially modulated phase factors $e^{iq_0x}$ and $e^{-iq_0x}$ in the upper and lower superconducting layers. **c**, HAADF image of $Ba_6Ta_{11}S_{28}$ superlattice taken along the [100] zone axis (scale bar: 1 nm) superimposed by a schematic of lattice structure. Ba, Ta and S atoms are painted with pink, purple, and yellow spheres, respectively. **d**, The atomic-resolved STM topography of the H-$TaS_2$ terminated $Ba_6Ta_{11}S_{28}$ superlattice ($15 \times 15$ nm$^2$, the sample bias $V_s = 0.05$ V and the tunneling current $I_s = 1.0$ nA). **e**, Temperature dependence of resistivity $\rho_{ab}(T)$ in $Ba_6Ta_{11}S_{28}$ (sample S1). The bottom inset shows the zoom in of the superconducting transition,

where $T_c^{onset} = 2.72$ K and $T_c^{zero} = 2.62$ K.

**The orbital effect induced finite-momentum pairing state in a bulk superconducting superlattice**

Figure 2 summarizes the magnetic field response of superconductivity in $Ba_6Ta_{11}S_{28}$ crystal. The temperature dependence of resistivity $\rho_{ab}(T)$ under out-of-plane and in-plane magnetic fields are plotted in Fig. 2a and 2b, respectively. The superconductivity is susceptible under out-of-plane fields and almost destroyed under 3 T (Fig. 2a). In contrast, the superconductivity is remarkably robust under in-plane magnetic fields (Fig. 2b). Color rendering of normalized resistivity ($\rho/\rho_n$) is plotted as functions of temperature and in-plane magnetic field ($\theta = 90°$), where the white color in the phase diagram represents approximately half of the normal-state resistivity and hence corresponds to the in-plane upper critical field $B_{c2,\parallel}(T)$ (Fig. 2c). Interestingly, a pronounced upturn behavior is revealed in $B_{c2,\parallel}(T)$ at low temperatures and high magnetic fields. To examine the evolution of the upturn behavior with field orientation, we measure the superconducting transition with the magnetic field applied along $\theta = 89.5°, 89°, 88°$ and $85°$ (Fig. 2c). The enhanced critical field as well as the upturn behavior is significantly suppressed even when the field is tilted $0.5°$ from the in-plane direction. At $\theta = 85°$, the upturn behavior almost disappears and $B_{c2}(T)$ exhibits a quasi-linear dependence. Figure 2d summarizes $B_{c2}(T)$ at different field angles, confirming that the upturn behavior in $B_{c2}(T)$ with an ultra-large critical field and the associated exotic superconducting states are linked to the in-plane orientation of the magnetic field.

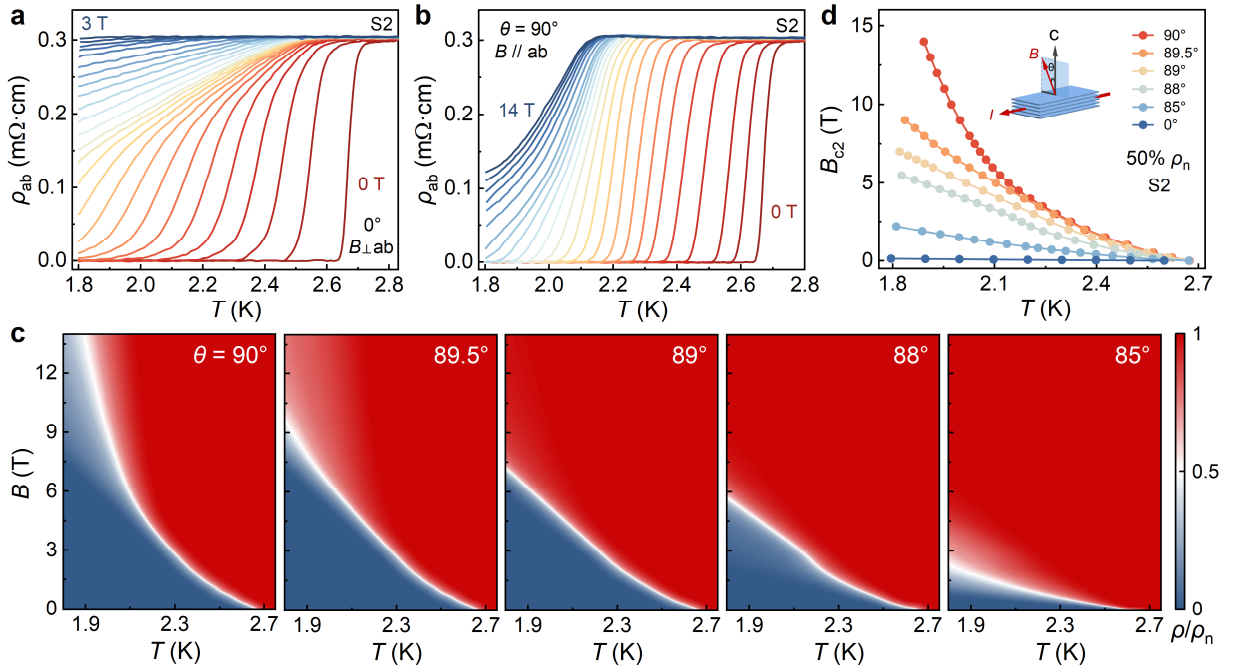

**Fig. 2 | The upturn behavior in temperature dependent in-plane critical field for $Ba_6Ta_{11}S_{28}$ (sample S2). a,b,** Temperature dependence of the resistivity $\rho_{ab}(T)$ under out-of-plane (**a**) and

in-plane (**b**) magnetic fields. The out-of-plane magnetic fields are varying in 0.01 T steps from 0 to 0.13 T, 0.15 T, 0.18 T, 0.22 T, 0.28 T, 0.35 T, 0.45 T, 0.6 T, 0.9 T, 1.2 T, 1.5 T, 2 T and 3 T. The in-plane magnetic fields are 0 T, 0.2 T, varying in 0.5 T steps from 0.5 T to 8 T, and in 1 T steps from 9 T to 14 T. **c**, Color rendering of normalized resistivity ($\rho/\rho_n$) as functions of temperature and field at different angles ($\theta$, defined in the inset of Fig. 2d) close to the in-plane orientation. With decreasing temperature, the in-plane critical field shows a pronounced upturn behavior, which is significantly suppressed when the field angle ($\theta$) tilts away from 90°. **d**, The critical fields versus temperature at different angles. The critical fields are determined by the criterion of 50% normal-state resistivity.

To investigate the exotic superconducting state of $Ba_6Ta_{11}S_{28}$ superlattice at higher fields and lower temperatures, we perform transport measurements under pulsed high magnetic field up to 41 T. Figure 3a shows the in-plane field dependence of resistivity $\rho_{ab}(B)$ at temperatures ranging from 1.06 to 2.76 K. The $\rho_{ab}(B)$ curves exhibit a two-step superconducting transition feature at low temperatures. Specifically, $\rho_{ab}$ increases quickly from zero to a resistive state near the characteristic field $B^*$. Above $B^*$, $\rho_{ab}$ increases slowly with rising field and is still below half of the normal state resistivity under a high in-plane field of 41 T at 1.06 K, yielding an exceedingly high $B_{c2,\parallel}$ beyond eight times the Pauli limit $B_P$. Here $B_P = 5.0$ T for $Ba_6Ta_{11}S_{28}$ superlattice. As shown in Fig. 3b, $B_{c2,\parallel}(T)$ starts to deviate from the linear dependence at 2.3 K and exhibits a pronounced upturn behavior at lower temperatures. These unique features shown in pulsed high field measurements have also been confirmed in a hybrid magnet with steady high magnetic field up to 44 T (Supplementary Fig. S3).

The anomalous upturn of $B_{c2,\parallel}(T)$ can be attributed to the orbital effect induced by in-plane magnetic field in the layered superconductor $Ba_6Ta_{11}S_{28}$ crystal. A large spin splitting of 120 meV is revealed in the $Ba_6Ta_{11}S_{28}$ superlattice by ARPES measurements[25], thus the spin pair-breaking effect is significantly suppressed due to the large Ising SOC and the orbital effect plays an essential role. The in-plane magnetic field induced orbital effect gives rise to the spatially modulated superconducting state $\Delta(x)e^{iq_lx}$(here $x$ denotes the in-plane direction and is perpendicular to the magnetic field), which can be well captured by a generalized Lawrence-Doniach (LD) model (see Supplementary Materials). The $B_{c2,\parallel}(T)$ from our theoretical model (crimson solid curve in Fig. 3b) is in a good agreement with the experimental phase boundary (red dot in Fig. 3b). The Cooper pair momentum $q_l$ in adjacent superconducting layers differs by a constant value $2q_0 = 2eBD/\hbar$ (where $D$ is the spacing between adjacent layers), and the amplitude of the order parameter $\Delta(x)$ oscillates in the intralayer direction with the period of $\pi/q_0$ (see Supplementary Materials for details), analogous to the oscillation in the LO state[14].

We further investigate the angular dependence of the finite-momentum pairing state (Supplementary Fig. S4) in a high-resolution sample rotator with an inductive angle sensor. The angle-dependent critical field $B_{c2}(\theta)$ at 1.82 K is summarized in Fig. 3c, which shows a sharp cusp near the in-plane orientation. Within a small tilted angle $\Delta\theta = |\theta - 90°| \leq 0.5°$, a sharp enhancement of critical field follows the trend given by the 2D Tinkham model[18] and consistent with the pronounced upturn behavior within $\Delta\theta \leq 0.5°$ (Fig. 2c). For $\Delta\theta > 0.5°$, the critical field varies smoothly with tilting angle and can be fitted using the 3D anisotropic Ginzburg-Landau (GL) model[18]. When the magnetic field is tilted away from the in-plane direction ($\Delta\theta > 0.5°$), the perpendicular component of the field leads to pancake vortices in the layered superconductor $Ba_6Ta_{11}S_{28}$, which significantly enhance the phase fluctuations and thus suppress the finite-momentum pairing state.

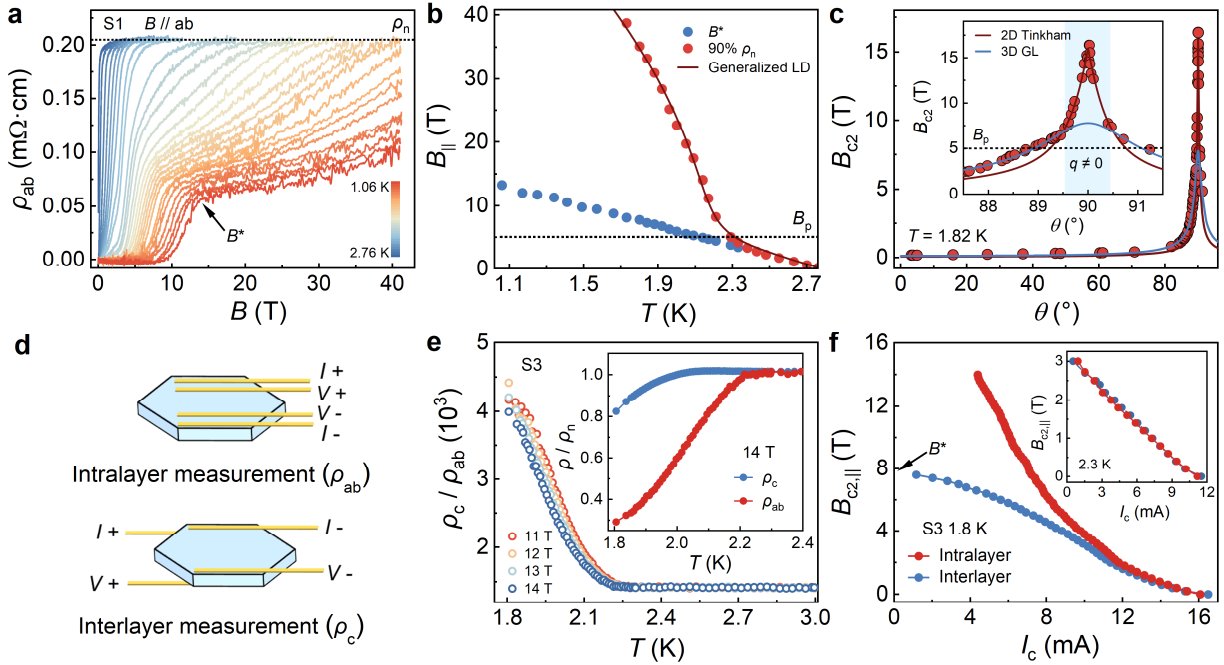

**Fig. 3 | The orbital effect induced finite-momentum pairing state and the suppression of interlayer coherence under high in-plane fields. a**, The in-plane field dependence of the resistivity $\rho_{ab}(B)$ at different temperatures (sample S1). **b**, The upper in-plane critical fields determined by 90% normal-state resistivity and the characteristic fields $B^*$. The crimson solid line is a theoretical fitting curve based on the generalized LD model. The dashed line denotes the Pauli limit $B_P = 1.86T_c$ (T/K) = 5.0 T for $Ba_6Ta_{11}S_{28}$. **c**, Angular dependence of the critical field $B_{c2}(\theta)$ measured at 1.82 K. The inset shows the enlarged view near 90°. Within a small angle $\Delta\theta = |\theta - 90°| < 0.5°$, $B_{c2}(\theta)$ is fitted using the 2D Tinkham model. For $\Delta\theta > 0.5°$, $B_{c2}(\theta)$ can be fitted using 3D anisotropic GL model. **d**, The schematic for the intralayer resistivity $\rho_{ab}$ (top) and interlayer resistivity $\rho_c$ (bottom) measurements. **e**, The ratio $\rho_c/\rho_{ab}$ versus temperature at different in-plane fields. The inset shows the temperature dependent normalized resistivity $\rho/\rho_n$ along the ab plane and the c axis at 14 T (sample S3). **f**, The in-plane critical fields with respect to the intralayer and interlayer critical currents at 1.8 K. The characteristic field $B^*$ denotes the field above which the interlayer coherence is suppressed. The inset shows the critical currents at 2.3 K.

**Suppression of interlayer coherence under high in-plane fields**

To investigate the two-step superconducting transition characterized by the field $B^*$, we conduct simultaneous measurements of the intralayer resistivity $\rho_{ab}$ (with current $I \; // \; ab$) and the interlayer resistivity $\rho_c$ ($I \; // \; c$), as depicted in the schematic of Fig. 3d. Surprisingly, $\rho_{ab}$ and $\rho_c$ exhibit distinct responses at low temperatures and under high in-plane fields. Pronounced upturn behavior is observed in $B_{c2,\|}(T)$ for $\rho_{ab}$ measurement, while the upturn behavior is not obvious in $B_{c2,\|}(T)$ for $\rho_c$ (Supplementary Fig. S5). Consistently, at an in-plane field of 14 T, $\rho_{ab}$ begins to decrease sharply at 2.2 K, whereas $\rho_c$ remains unchanged until 2 K (the inset of Fig. 3e). Figure 3e also illustrates the anisotropy between $\rho_{ab}$ and $\rho_c$. The enhanced $\rho_c/\rho_{ab}$ ratio under large magnetic field indicates the suppression of coherent coupling between the superconducting layers.

This suppression of interlayer coherence, evident when the field exceeds $B^*$, is further corroborated by the significant anisotropy observed between the superconducting critical currents in the interlayer ($I \; // \; c$) and intralayer ($I \; // \; ab$) directions. At 1.8 K, there is a noticeable upturn in the $B_{c2,\|}$ versus intralayer critical current, while the $B_{c2,\|}$ exhibits a saturating trend with respect to interlayer critical current (Fig. 3f and Supplementary Fig. S6). The difference between the intralayer and interlayer critical current gradually weakens with increasing temperature (Supplementary Fig. S7) and almost disappears at 2.3 K (the inset of Fig. 3f). In $Ba_6Ta_{11}S_{28}$ superlattice, the suppression of interlayer coherence under high in-plane magnetic field ($B > B^*$) may originate from the melting of Josephson vortex solid[27-29]. Following the melting transition, the in-plane movement of Josephson vortices suppresses the interlayer coherence and enhances the phase fluctuations[30,31] of the superconducting H-$TaS_2$ layers, which may lead to the finite resistivity of the sample. According to our theoretical model, the melting of Josephson vortex solid gives rise to a uniform in-plane magnetic field, which is the prerequisite of the orbital effect induced finite-momentum pairing state. As a result, the spatial modulation of this finite-momentum pairing state becomes pronounced across the melting line (Supplementary Fig. S8), demonstrating the crucial role of Josephson vortex solid melting in the spatial modulation.

**Phase diagram and disorder effect**

Figure 4a presents the phase diagram of the spatially modulated superconducting state in $Ba_6Ta_{11}S_{28}$ superlattice, induced by in-plane magnetic field. The upper critical field with the pronounced upturn feature characterizes the pairing-breaking process, while the characteristic field $B^*$ marks the suppression of coherent coupling between the superconducting layers. The suppressed interlayer coherence potentially stems from the melting transition of the Josephson vortex solid (the schematic shown in the left of Fig. 4a), evidenced by the good agreement between the characteristic fields $B^*$ and the calculated melting line (see Fig. 4a and Supplementary Materials for detailed discussions). The Josephson vortex solid melting leads to highly anisotropic transport features observed in Figs. 3e-f.

Commonly, clean superconducting system is a prerequisite for the FFLO state[15,32]. Nevertheless, the observation of orbital effect induced finite-momentum pairing state in $Ba_6Ta_{11}S_{28}$ is in the dirty limit, which is characterized by $\ell < \xi_0$ (where the mean free path $\ell = 7.8$ nm and the Pippard coherence length $\xi_0 = 211$ nm, see Methods for details). The strong disorder in the $Ba_3TaS_5$ block layer is further evidenced by HAADF and SAED images (Fig. 1c and Supplementary Fig. S1). To examine the disorder effect on the finite-momentum pairing state, we measure the in-plane critical field $B_{c2,\parallel}(T)$ of $Ba_6Ta_{11}S_{28}$ crystals with different disorder strength. The strength of disorder can be characterized by the normal-state resistivity $\rho_{ab}$ (3 K) and RRR, where larger $\rho_{ab}$ (3 K) and smaller RRR indicate higher disorder strength (Supplementary Table S1). As shown in Fig. 4b, the upturn behavior becomes weaker in the samples with higher disorder strength. The spin-orbit scattering described by Klemm-Luther-Beasley (KLB) model[33] can be ruled out since stronger disorder usually give rises to larger $B_{c2,\parallel}$ (see Supplementary Materials for detailed discussions). Our results demonstrate that the orbital effect induced finite-momentum pairing states are robust to moderate disorder strength. When further increasing the disorder level, the phase fluctuation becomes pronounced and suppresses the spatial modulation of the finite-momentum pairing states.

Finally, we compare the in-plane critical field $B_{c2,\parallel}(T)$ of $Ba_6Ta_{11}S_{28}$ superlattice with representative superconducting systems in the previous literatures (Figs. 4c and 4d). As presented in Fig. 4d, large in-plane critical field is reported in atomically thin TMD materials showing Ising superconductivity[1-4] but significantly weakened in thicker films and bulk TMDs[1,4,26,34,35]. Remarkably, for $Ba_6Ta_{11}S_{28}$ superlattice, the extrapolated $B_{c2,\parallel}(0) \approx 12B_P$ (Fig. 4c), much higher than that of bulk 2H-$TaS_2$ and even comparable to that of monolayer Ising superconductors. Besides, the $B_{c2,\parallel}(T)$ of intercalated $TaS_2$ compounds (e.g., 2H-$TaS_2$+(PY)$_{0.5}$[36] and 4Hb-$TaS_2$[37]) are plotted for comparison, where no obvious upturn behavior is observed. The exceedingly large in-plane critical field in the orbital effect induced finite-momentum pairing state of $Ba_6Ta_{11}S_{28}$ demonstrates the intriguing interplay between superconductivity, strong SOC and orbital effect, thereby providing a fertile playground to explore nonreciprocal transport as well as unconventional and topological superconductivity[38-42]. Our findings on the finite-momentum pairing state in the superlattice material composed of sequent superconducting and block layers may inspire further investigations on the exotic modulated superconducting state in strongly correlated systems, notably in high-$T_c$ superconductors[43-47].

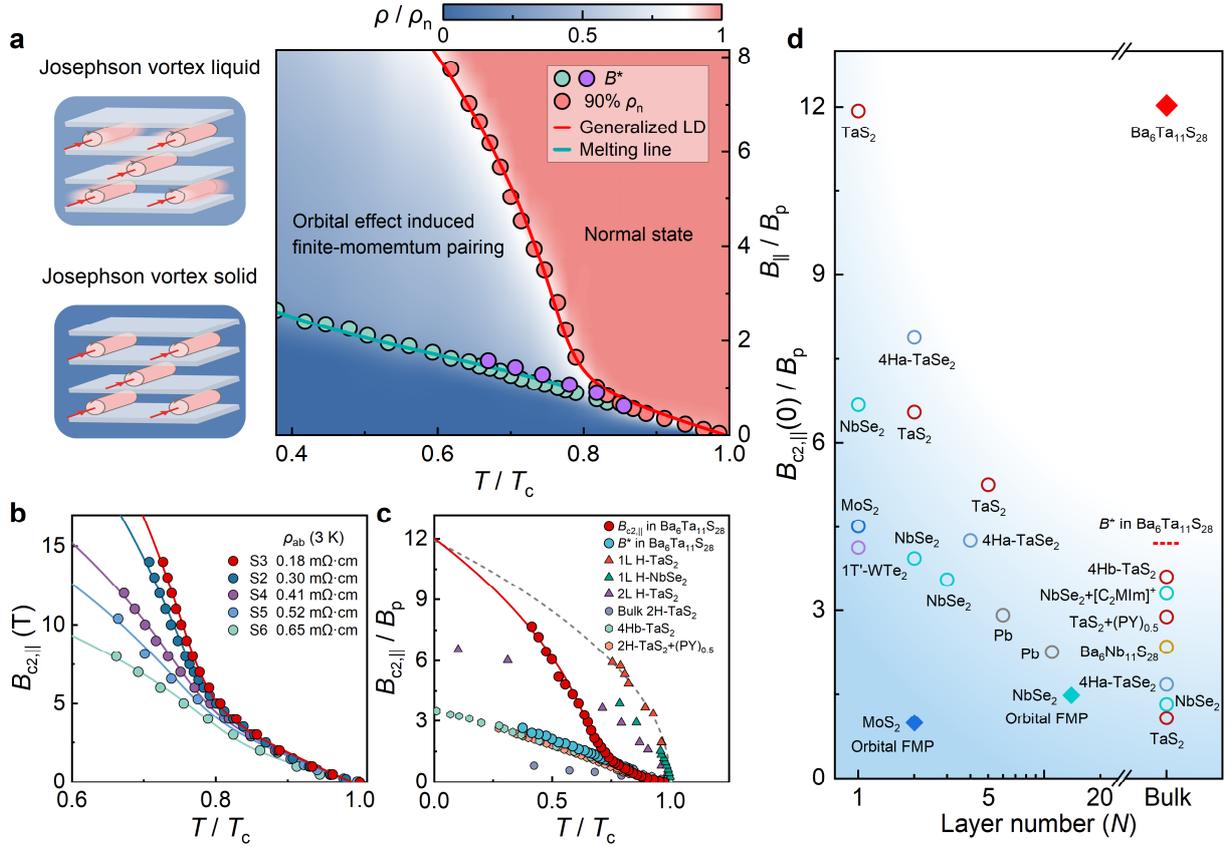

**Fig. 4 | The $B$ versus $T$ phase diagram for Ba$_6$Ta$_{11}$S$_{28}$ superlattice. a**, The superconducting phase diagram is depicted by the color rendering of $\rho/\rho_n$ as functions of temperature and in-plane magnetic field. The red solid line is the fitting curve of upper critical field based on the generalized LD model, and the dark cyan solid line is the fitting curve of the characteristic fields $B^*$ based on the Josephson vortex solid melting. A schematic of the melting transition from the Josephson vortex solid to the Josephson vortex liquid is plotted on the left of the phase diagram. The solid cyan circles are defined in Fig. 3a and the solid purple circles are defined in Fig. 3f. **b**, The in-plane critical field versus temperature in samples with different disorder strengths. The pronounced upturn behavior becomes weaker in the samples with higher disorder strength characterized by larger normal-state resistivity at 3 K. **c**, The normalized in-plane critical field $B_{c2,\parallel}/B_p$ as a function of reduced temperature $T/T_c$ for representative TMD materials[1,4,26,36,37]. Bulk Ba$_6$Ta$_{11}$S$_{28}$ exhibits a very large in-plane critical field which is even comparable to monolayer Ising superconductors at low temperatures. The gray dashed line is the fitting curve based on 2D GL formula $B_{c2,\parallel} = B_{c2,\parallel}(0)\sqrt{1-T/T_{c0}}$ for monolayer 1H-TaS$_2$[4]. The solid lines in **b** and **c** are the fitting curves based on the generalized LD model. **d**, Thickness dependence of the normalized in-plane critical field $B_{c2,\parallel}/B_p$ at 0 K in representative superconducting systems[1,3,4,9,22,23,26,35-37,48,49]. Generally, the in-plane critical field decreases with increasing thickness. The red solid diamond represents the in-plane critical field of Ba$_6$Ta$_{11}$S$_{28}$ superlattice, which is much higher than

that of bulk 2H-TaS$_2$ and even comparable to that of monolayer Ising superconductors. The solid diamonds represent the in-plane critical fields of orbital effect induced finite-momentum pairing states (orbital FMP) in different materials. Unless otherwise noted, the TMD materials in this panel are 2H phase.

**Methods**
**Single crystal synthesis and characterization**
High-quality Ba$_6$Ta$_{11}$S$_{28}$ single crystals were grown by the flux method starting from a mixture of BaS, Ta, S, and BaCl$_2$. The mixture was sealed into quartz ampoules after pumping down to less than 10$^{-4}$ Pa. Subsequently, the ampoule was loaded into a box furnace, which was heated up to 950°C at 50 °C/h and held at 950°C for 120 hours. The temperature was then lowered to 750°C at 2°C/h followed by furnace cooling. Eventually, lustrous black thin plates with typical hexagon were extracted. Energy dispersive X-ray spectroscopy (EDX, Oxford X-Max 50) was performed to determine the chemical composition. The atomic proportion was consistent with the composition of Ba$_6$Ta$_{11}$S$_{28}$ within the instrumental resolution.

**Scanning transmission electron microscopy (STEM)**
Atomic-resolution high-angle annular dark field scanning transmission electron microscopy (HAADF-STEM) images were obtained with a spherical-aberration corrected JEOL ARM200F instrument equipped with double aberration correctors and operated at 200 kV. The transmission electron microscope (TEM) sample along the [100] zone-axis direction was prepared through a focused ion beam (FIB), while [001] zone-axis samples were made by mechanical exfoliation method with scotch tape.

**Scanning tunneling microscopy (STM) characterization**
The STM measurements were performed in an Omicron STM instrument at 4.3 K. The in-situ cleavage of Ba$_6$Ta$_{11}$S$_{28}$ single crystals was conducted in an ultrahigh vacuum chamber.

**Transport measurements and diamagnetic measurements**
The resistance and magnetoresistance were measured in a commercial physical property measurement system (Quantum Design, PPMS-14). Standard four-probe method (Fig. 3d in the main text) was used for the intralayer transport measurements ($\rho_{ab}$), and Montgomery method (Fig. 3d in the main text) was used for interlayer transport measurement ($\rho_c$). To ensure the same measurement conditions, $\rho_{ab}$ and $\rho_c$ were measured simultaneously on the same sample. For angular dependent magnetoresistivity measurements, the samples were mounted on a rotation holder with an angle precision better than 0.1°. The *I-V* characteristics were performed in PPMS-14 using electrical transport option.

The pulsed high magnetic field measurements were carried out in a He-3 fridge with pulsed magnetic field up to 41 T in Wuhan National High Magnetic Field Center. The angular dependent transport measurement was based on a high-precision sample rotation rod, where the tilted angle was determined by an inductive coil providing an angle accuracy better than 0.1°.

The steady high magnetic field measurements were performed in a He-3 refrigerator in the High Magnetic Field Laboratory at Hefei. The magnetic field range is 10-44 T for the hybrid magnet, which consists of a superconducting outsert of 10 T and a water-cooled insert of 0-34 T. The sample was mounted on the rotation holder for the measurement under parallel magnetic field.

We have also performed the diamagnetic measurements, which was carried out in a magnetic property measurement system (Quantum Design, MPMS-3).

**Data Availability**

All data that support the findings of this study are available from the corresponding author on reasonable request.

**Code Availability**

The code used to analyze the data reported in this study is available from the corresponding authors on reasonable request.

**Acknowledgments:** We thank Peter Fulde, K. T. Law, Ziqiang Wang, Chaoxing Liu, Haoran Ji for fruitful discussions. We thank Gangjian Jin, Senyang Pan, Yuhao Ye for the help in high magnetic field measurements. We thank Wei Ren for the help in scanning tunneling microscopy measurements. This work was financially supported by the National Natural Science Foundation of China (Grant No. 12488201 (J.W.)), the National Key Research and Development Program of China (Grant No. 2024YFA1409002 (T.-L.X.), No. 2022YFA1403103 (Yi Liu), No. 2023YFA1406500 (Yi Liu), No. 2019YFA0308602 (T.-L.X.), No. 2024YFA1409001 (H.L.)), the Innovation Program for Quantum Science and Technology (2021ZD0302403 (J.W.)), the National Natural Science Foundation of China (No. 12174442 (Yi Liu), No. 12074425 (T.-L.X.), No. 12374037 (H.L.)), Young Elite Scientists Sponsorship Program by CAST (No. 2023QNRC001 (Yi Liu)), Young Elite Scientists Sponsorship Program by BAST (No. BYESS2023452 (Yi Liu)), Guangdong Provincial Quantum Science Strategic Initiative (GDZX2401001 (Yanzhao Liu), GDZX2401009 (Yanzhao Liu)), the Fundamental Research Funds for the Central Universities, and the Research Funds of Renmin University of China (No. 23XNKJ22 (T.-L.X.)). We thank the Hybrid Magnet (31125.02.SHMFF.HM) at the Steady High Magnetic Field Facility, CAS, for providing technical support and assistance in data collection.

**Author contributions:** J.W., Yi Liu and T.-L.X. conceived and supervised the project. J. Lin, Y.Z. synthesized and characterized the single crystals under the guidance of T.-L.X.; J. Lin, Z.W., Z.C., R.L., Yi Liu performed the transport measurements and analyzed the data under the guidance of J.W., with the contribution from H. Yan, H.L. and T.-L.X.; C.X., Z.Z. and H.Z. helped in high magnetic field measurements; X.M. and H. Yang performed scanning transmission electron microscopy measurements; Y. Lei, J. Liu, Yanzhao Liu performed scanning tunneling microscopy measurements under the guidance of J.W.; H. Yan and H.L. performed the theoretical analysis; J. Lin, Z.W., Yi Liu and J.W. wrote the manuscript with input from H. Yan, X.M., Yanzhao Liu, T.-L.X. and H.L. All authors contributed to the related discussions.

**Competing interests**
The authors declare no competing interests.

Supplementary Information for

# Exceedingly large in-plane critical field of finite-momentum pairing state in bulk superlattices

**Contents**

1. Angular dependence of the upper critical field $B_{c2}(\theta)$
2. Theoretical model for the orbital effect induced finite-momentum pairing state
3. Josephson vortex lattice melting
4. Characterization of the dirty limit
5. Discussions on the difference between the FFLO state and orbital effect induced finite-momentum paring state
6. Discussions on other possible scenarios for the upturn behavior in $B_{c2}(T)$

Supplementary References

# 1. Angular dependence of the upper critical field $B_{c2}(\theta)$

For bulk superconductors, the angular dependence of the upper critical field $B_{c2}(\theta)$ can be described by the 3D anisotropic Ginzburg-Landau (GL) formula[1]

$$\left(\frac{B_{c2}(\theta)\sin\theta}{B_{c2,\parallel}}\right)^2 + \left(\frac{B_{c2}(\theta)\cos\theta}{B_{c2,\perp}}\right)^2 = 1 \qquad (S1)$$

Here, $\theta$ is the angle between the c axis and the applied magnetic field, $B_{c2,\parallel}$ and $B_{c2,\perp}$ are the upper critical fields with the field parallel ($\theta = 90°$) and perpendicular ($\theta = 0°$) to the ab plane, respectively.

For two-dimensional superconductors, $B_{c2}(\theta)$ is typically described using the 2D Tinkham model which takes the form[1]

$$\left(\frac{B_{c2}(\theta)\sin\theta}{B_{c2,\parallel}}\right)^2 + \left|\frac{B_{c2}(\theta)\cos\theta}{B_{c2,\perp}}\right| = 1 \qquad (S2)$$

The 3D anisotropic GL formula exhibits a dome shape near the in-plane direction, while the 2D Tinkham model shows a sharp cusp. In our work, a combination of 3D anisotropic GL and 2D Tinkham model is used to fit the two-stage angular dependence of the upper critical field $B_{c2}(\theta)$ at 1.82 K (Fig. 3c in the main text). Within a small tilted angle $\Delta\theta = |\theta - 90°| \leq 0.5°$, a sharp enhancement of critical field follows the trend given by the 2D Tinkham model. For $\Delta\theta > 0.5°$, the critical field varies smoothly with tilting angles and can be fitted using the 3D anisotropic GL model. At 2.4 K, $B_{c2}(\theta)$ can be well described by the 3D GL model for the entire angular regime (Fig. S9).

## 2. Theoretical model for the orbital effect induced finite-momentum pairing state

The superlattice $Ba_6Ta_{11}S_{28}$ is comprised of alternating superconducting layers of 1H-$TaS_2$ and block layers of $Ba_3TaS_5$. The adjacent superconducting layers are coupled via the Josephson coupling. Due to the strong Ising spin-orbit coupling (SOC) in 1H-$TaS_2$ layers, the electron spins are locked in the out-of-plane direction, which significantly suppress the Zeeman effect of the in-plane magnetic field. Therefore, the Zeeman pair-breaking effect is ignorable and the orbital effect plays an essential role under the in-plane magnetic field. To describe the temperature dependence of in-plane critical field in $Ba_6Ta_{11}S_{28}$ superlattice, we consider the generalized Lawrence-Doniach (LD) model [2-4]. The free energy consists of the energy of each individual layer and the Josephson coupling between the adjacent superconducting layers.

$$F = \sum_l \int d^2 r_\| \{ \int_{lD-\frac{d}{2}}^{lD+\frac{d}{2}} dz \, [\frac{\hbar^2}{2M_\|} \left| \left( \nabla_\| - i\frac{2\pi}{\Phi_0} A_\| \right) \Psi_l(r_\|, z) \right|^2 + \frac{\hbar^2}{2M_\perp} \left| \left( \nabla_\perp - i\frac{2\pi}{\Phi_0} A_\perp \right) \Psi_l(r_\|, z) \right|^2$$

$$+ \alpha |\Psi_l(r_\|, z)|^2 + \frac{\beta}{2} |\Psi_l(r_\|, z)|^4] - \frac{1}{2} Jd [\Psi_l^* \left( r_\|, lD + \frac{d}{2} \right) \Psi_{l+1} \left( r_\|, (l+1)D - \frac{d}{2} \right) e^{-i\varphi_{l,l+1}}$$

$$+ \Psi_l^* \left( r_\|, lD - \frac{d}{2} \right) \Psi_{l-1} \left( r_\|, (l-1)D + \frac{d}{2} \right) e^{-i\varphi_{l-1,l}} + c.c.]\}, \quad (S1)$$

$$\text{with } \varphi_{l,l+1}(r_\|, z) = \frac{2\pi}{\Phi_0} \int_{lD+d/2}^{(l+1)D-d/2} dz \, A_\perp(r_\|, z). \quad (S2)$$

In Eq. S1, $\Phi_0 = h/2e$ and we assume that the superconducting layers possess a finite thickness d and an interlayer spacing D. The Josephson coupling strength, represented by J, is typically small in our system. Here, $M_\|$ and $M_\perp$ denote the effective masses in the in-plane and out-of-plane directions, respectively. The order parameter for the $l$-th layer is denoted as $\Psi_l(r_\|, z)$, while $\alpha$ and $\beta/2$ are the coefficients associated with the second-order and fourth-order terms, where $\alpha = \alpha_0(T - T_{c0})$, and $T_{c0}$ signifies the transition temperature of an individual layer. In Eq. S2, $\varphi_{l,l+1}$ denotes the Peierls phase, which ensures gauge invariance in the presence of a magnetic field.

Suppose the external magnetic field $B = B(0, \sin\theta, \cos\theta)$, where $\theta$ is the angle between the magnetic field and the $z$ axis. By choosing the gauge $A = B(z\sin\theta, x\cos\theta, 0)$, Eq. S1 transforms to

$$F = \sum_l \int d^2 r_\| \{ \int_{lD-\frac{d}{2}}^{lD+\frac{d}{2}} dz \, [\frac{\hbar^2}{2M_\|} \left| \left( \frac{\partial}{\partial x} - i\frac{2\pi}{\Phi_0} B\sin(\theta)z \right) \Psi_l(r_\|, z) \right|^2$$

$$+ \frac{\hbar^2}{2M_\|} \left| \left( \frac{\partial}{\partial y} - i\frac{2\pi}{\Phi_0} B\cos(\theta)x \right) \Psi_l(r_\|, z) \right|^2 + \frac{\hbar^2}{2M_\perp} \left| \frac{\partial}{\partial z} \Psi_l(r_\|, z) \right|^2 + \alpha |\Psi_l(r_\|, z)|^2 + \frac{\beta}{2} |\Psi_l(r_\|, z)|^4]$$

$$- \frac{1}{2} Jd [\Psi_l^* \left( r_\|, lD + \frac{d}{2} \right) \Psi_{l+1} \left( r_\|, (l+1)D - \frac{d}{2} \right) e^{-i\varphi_{l,l+1}}$$

$$+ \Psi_l^* \left( r_\|, lD - \frac{d}{2} \right) \Psi_{l-1} \left( r_\|, (l-1)D + \frac{d}{2} \right) e^{-i\varphi_{l-1,l}} + c.c.]\}. \quad (S3)$$

Given $A_z = 0$ in our gauge selection, the Peierls phase $\varphi_{l,l+1}$ equals zero. $\Psi_l(r_\|, z)$ in Eq. S3 should be independent on $z$ to minimize the free energy.

In the context of bulk materials, we can safely neglect the boundary effects, allowing us to represent the amplitude of the order parameter for all layers with a single function $\Delta$. Thus, the order parameter takes the form $\Psi_l(x) = \Delta(x)e^{i\varphi_l(x)}$. Assuming the phase $\varphi_l(x) = Q_l x$, where $Q_l$ denotes the mechanical momentum of the Cooper pairs in the $l$-th layer. We find that as the layer number approaches infinity, the in-plane kinetic energy can be very large if $Q_l$ remains constant across all layers (the uniform superconducting state is a special case for $Q_l = 0$). Based on this analysis, we propose that the mechanical momentum should be layer-dependent and independent of z coordinate within an individual superconducting layer. Specifically, the order parameter can be expressed as $\Psi_l(x) = \Delta(x)e^{iQ_l x}$, with $Q_l = 2\pi B\sin(\theta)lD/\Phi_0$ and $l$ is an integer denoting the $l$-th layer. The momentum of the adjacent layers varies by a constant value $2q_0 = 2\pi B\sin(\theta)D/\Phi_0$, which is proportional to the external magnetic field.

Thus, the free energy then transforms to

$$F = d\sum_l \int d^2 r_\parallel \left\{ \frac{\hbar^2}{2M_\parallel}\left[\left|\frac{d}{dx}\Delta(x)\right|^2 + \frac{d^2}{12}\left(\frac{2\pi}{\Phi_0}B\sin(\theta)\right)^2 |\Delta(x)|^2 + \left(\frac{2\pi}{\Phi_0}B\cos(\theta)x\right)^2 |\Delta(x)|^2\right] + \alpha|\Delta(x)|^2 + \right.$$

$$\left. \frac{\beta}{2}|\Delta(x)|^4 - 2J\cos\left(\frac{2\pi}{\Phi_0}B\sin(\theta)Dx\right)|\Delta(x)|^2 \right\}. \tag{S4}$$

To render the free energy dimensionless, we introduce the x coordinate scale $L_0$, temperature scale $T_0$, and magnetic field scale $B_0$, where $L_0 = \Phi_0/\pi B\sin(\theta)D$, $T_0 = J/\alpha_0$, and $B_0 = \Phi_0/2\pi\lambda_J D$, where $\lambda_J = \sqrt{\hbar^2/2M_\parallel J}$. In the absence of an external magnetic field, as the temperature approaches the superconducting transition temperature $T_c$ of the system, the spatial derivative of the amplitude $d\Delta(x)/dx$ approaches zero more rapidly than the amplitude $\Delta(x)$ itself. By neglecting the $d\Delta(x)/dx$ term and the fourth-order term in Eq. S4, and varying the free energy with respect to $\Delta(x)$, we arrive at $\alpha(T=T_c) - 2J = 0$ and the relation between $T_{c0}$ and $T_c$, $T_c = T_{c0} + 2T_0$. This result implies that for negative Josephson coupling, the transition temperature of the bulk system is greater than that of a single layer by a temperature $2T_0$.

Next, we introduce the rescaled coordinate $\bar{x} = x/L_0$, rescaled temperature $t = (T - T_c)/T_0$ and rescaled magnetic field $b = B\sin(\theta)/B_0$. Consequently, Eq. S4 transforms into

$$F = JL_y d \sum_l \int d\bar{x} \left\{ \frac{b^2}{4}\left|\frac{d}{d\bar{x}}\Delta(\bar{x})\right|^2 + \frac{b^2}{12}\left(\frac{d}{D}\right)^2 |\Delta(\bar{x})|^2 + b^2\left(\frac{\Phi_0}{\pi BD^2}\frac{\cos(\theta)}{\sin^2(\theta)}\right)^2 \bar{x}^2|\Delta(\bar{x})|^2 \right.$$

$$\left. +(t+2)|\Delta(\bar{x})|^2 + \frac{\beta}{2J}|\Delta(\bar{x})|^4 - 2\cos(2\bar{x})|\Delta(\bar{x})|^2 \right\}. \tag{S5}$$

We then vary Eq. S5 with respect to the $\Delta(\bar{x})$, and neglect the fourth-order term, obtaining the equation that corresponds to the condition for minimum free energy.

$$\left[\frac{d^2}{d\bar{x}^2} - \frac{1}{3}\left(\frac{d}{D}\right)^2 - \frac{4}{b^2}(t+2) + \frac{8}{b^2}\cos(2\bar{x}) - \left(\frac{2\Phi_0}{\pi BD^2}\frac{\cos(\theta)}{\sin^2(\theta)}\right)^2 \bar{x}^2\right]\Delta(\bar{x}) = 0 \quad (S6)$$

Eq. S6 can be converted into an eigenvalue equation:

$$\hat{L}\Delta(\bar{x}) = E\Delta(\bar{x})$$

$$\text{with } E = \frac{1}{3}\left(\frac{d}{D}\right)^2 + \frac{4}{b^2}(t+2),$$

$$\hat{L} = \frac{d^2}{d\bar{x}^2} + \frac{8}{b^2}\cos(2\bar{x}) - \left(\frac{2\Phi_0}{\pi BD^2}\frac{\cos(\theta)}{\sin^2(\theta)}\right)^2 \bar{x}^2. \quad (S7)$$

The cosine term requires that the amplitude of the order parameter has a periodic spatial modulation. We calculate this equation by differentiating over one period, using a sufficiently dense partition to enhance accuracy. We then diagonalize the matrix to find the eigenvalues. The upper critical field is determined by the largest eigenvalue of Eq. S7.

In the main text, we fit the experimental data by adjusting three key parameters (Figs. 3b and 4): the characteristic magnetic field scale $B_0$ corresponding to the Josephson vortex solid, the characteristic temperature scale $T_0$ corresponding to the interlayer Josephson coupling and the ratio $d/D$. The fitting parameters for samples S2-S6 are summarized in Table S2. Notably, the model can well describe the upturn behavior observed in samples across a wide range of disorder levels. In samples characterized by heightened levels of disorder, the fitted ratio $d/D$ is large. Large $d/D$ indicates a broadening of the superconducting region in the $z$ direction, allowing electrons to rotate in the $x - z$ plane, which in turn suppresses the upper critical field.

The amplitude of the order parameter for the finite-momentum pairing state can be determined by minimizing the free energy in Eq. (S5), and its periodic form is illustrated in Fig. S8. Although the magnetic vector potential and the configuration of finite-momentum pairing state depend on the choice of specific gauge, the crucial aspect is that the amplitude of the order parameter shows periodic oscillations, which is gauge-invariant and observable.

## 3. Josephson vortex lattice melting

The cyan and purple points in the phase diagram (Fig. 4a) indicate a first-order phase transition driven by the melting of the Josephson vortex lattice[4]. This melting line can be determined by the Lindemann criterion, which states that the Josephson vortices vibrate due to thermal fluctuations. When the ratio of the mean square displacement to the square of the lattice constant $\langle u^2 \rangle / L_0^2$ exceeds a certain value $c_L^2$ (where $c_L$ is the Lindemann parameter), the lattice structure breaks down and results in the formation of a Josephson vortex liquid.

The mean square displacement of Josephson vortices can be calculated through[5],

$$\langle u^2 \rangle = k_B T \int_{-\infty}^{+\infty} \frac{dk_y}{\pi} \int_{BZ} \frac{d^2\boldsymbol{k}}{(2\pi)^2} \frac{1}{c_{11}k_x^2 + c_{44}k_y^2 + c_{66}\tilde{k}_z^2 + \alpha_L}. \quad (S8)$$

In the y direction that parallel to the magnetic field, periodicity is absent, allowing $k_y$ to take all the values. The integral in the x-z plane is performed over the Brillouin zone of the Josephson vortex lattice and $\tilde{k}_z = 2\sin(k_z D/2)/D$. Here, The Labusch parameter $\alpha_L$ denotes the restoring force exerted on the vortices per unit of their displacement. $c_{11}$, $c_{66}$ and $c_{44}$ represent the uniaxial compression modulus, shear modulus and tilting modulus, respectively. The expression of these moduli is written as follows[6]:

$$c_{11} = c_{44} = \frac{B^2 \epsilon}{4\pi \lambda_{ab}^2 \tilde{k}_z^2} |\bar{f}|^2, \quad c_{66} = \frac{\Phi_0^2 \epsilon}{32\pi^3 \gamma^4 \lambda_{ab}^2 D^2} |\bar{f}|^2. \tag{S9}$$

Here, $\epsilon = d/D$, $\lambda_{ab}$ is in-plane penetration length of the individual superconducting layers, satisfying $\lambda_{ab} = \lambda_{ab}(0)(1 - T/T_{c0})^{-1/2}$, where $T_{c0}$ is the superconducting transition temperature of a single layer. $\gamma$ denotes the anisotropy of the system. The reduced order parameter $f$, which characterizes the system in the presence of the dense Josephson vortex lattice, is defined as $f = \Psi/\Psi_0$, where $\Psi_0 = \sqrt{-\alpha/\beta}$. The anisotropy $\gamma$ and the in-plane coherence length of single layer $\xi_{ab}(0)$ can be determined from $(T_0, B_0)$ by using the relations $\gamma = \Phi_0/2\pi B_0 D^2$ and $\xi_{ab}(0) = \gamma D \sqrt{T_0/(T_c - 2T_0)}$.

The calculated melting line shows excellent agreement with the experimental data, as demonstrated in Fig. S8a. Furthermore, by minimizing the free energy before and after the melting process, we find that the spatial variation of the order parameter amplitude gets much more pronounced when $B > B^*$ (Fig. S8b), demonstrating the critical role of suppressed interlayer coherence in the spatial modulation.

## 4. Characterization of the dirty limit

The classification of superconductors in the clean or dirty limit is made by comparing the Pippard coherence length $\xi_0 = 0.18 \hbar v_F/k_B T_c$ with the normal state mean-free path $\ell = v_F \tau$, where $v_F$ is the Fermi velocity, $T_c$ is the superconducting transition temperature and $\tau$ is the normal state electron scattering time. Specifically, $\xi_0 \ll \ell$ for clean superconductors and $\ell < \xi_0$ for dirty superconductors. Our observation of the orbital effect induced finite-momentum pairing state is characterized by $\ell < \xi_0$, indicating the dirty limit.

To determine the mean free path $\ell$, we perform Hall measurements to obtain the carrier concentration $n$ and the mobility $\mu$. Figure S10a shows the Hall resistivity $\rho_{yx}$ as a function of magnetic field $B$ at different temperatures, which shows a linear dependence with respect to the field. $n$ and $\mu$ are related to the Hall coefficient $R_H$ via the relation $n = -\frac{1}{R_H e}$ and $\mu = \frac{1}{en\rho_{xx}}$ (Fig. S10b).

Using the Drude model, we estimate the relaxation time

$$\tau = \frac{m^*}{\rho n e^2} = 1.88 * 10^{-14} \text{ s} \qquad (S3)$$

Here, the effective mass $m^* \approx 0.76\ m_e$[7], and $\rho, n, e$ are the resistivity at 3 K (sample S3), the carrier concentration, and the elementary charge, respectively. Given the Fermi velocity

$$v_F = \left(\frac{\hbar}{m^*}\right) k_F = \left(\frac{\hbar}{m^*}\right)(3\pi^2 n)^{1/3} = 4.14 * 10^7 \text{ cm/s} \qquad (S4)$$

we evaluate the mean free path $\ell = v_F \tau = 7.8$ nm.

The band structure of $Ba_6Ta_{11}S_{28}$ along the $\Gamma - K$ direction is obtained by angle-resolved photoemission spectroscopy (ARPES)[7]. The quadratic fit of band structure yields the Fermi velocities, which are consistent with that estimated using the Drude model.
The Pippard coherence length of the superconducting state is estimated as $\xi_0 = 0.18\hbar v_F/k_B T_c = 211$ nm with $T_c = 2.7$ K, which yields $\xi_0/\ell = 27$ and indicates that $Ba_6Ta_{11}S_{28}$ is in the dirty limit.

The in-plane GL coherence length $\xi_{ab}$ at zero temperature can be extracted from the temperature dependence of the out-of-plane upper critical field using the GL model $B_{c2,\perp} = \frac{\Phi_0}{2\pi\xi_{ab}^2}(1 - T/T_c)$.

Here $\Phi_0 = \frac{h}{2e}$ is the flux quantum. As shown in Fig. S11 of the Supplementary Materials, the extracted $\xi_{ab}$ is 27.8 nm for sample S2 and 22.0 nm for sample S3. Hence, the mean free path is also smaller than the in-plane GL coherence length, $\ell < \xi_{ab}$, indicating the dirty limit.

## 5. Discussions on the difference between the FFLO state and orbital effect induced finite-momentum paring state

Commonly, Zeeman effect of external magnetic field gives rise to the mismatch between the spin-up and spin-down Fermi surface, which leads to the finite-momentum pairing state, namely the FFLO state[8,9]. According to previous investigations, the FFLO state appears at low temperatures with the critical field slightly beyond the Pauli limit[10-14]. Such Zeeman effect induced finite-momentum pairing state (i.e. FFLO state) is sensitive to the scattering from disorder and hence requires clean superconducting systems.

However, as we mentioned in the main text, the large Ising SOC in $Ba_6Ta_{11}S_{28}$ superlattice significantly suppresses the Zeeman effect of in-plane magnetic field, and thus the orbital effect plays an crucial role in the formation of finite-momentum pairing state. Moreover, in our work, the orbital effect induced finite-momentum pairing state shows exceedingly large in-plane critical field beyond 8 times the Pauli limit and appears in a board temperature regime of the phase diagram. Remarkably, the observed finite-momentum pairing state is robust to moderate disorder strength, which is distinct from the FFLO state that requires clean superconducting systems.

## 6. Discussions on other possible scenarios for the upturn behavior in $B_{c2}(T)$

An exceedingly large in-plane critical field $B_{c2,\parallel}$ and a pronounced upturn behavior in $B_{c2,\parallel}(T)$ are revealed as experimental indications of the orbital effect induced finite-momentum pairing state, where the interlayer coherence is suppressed under high in-plane fields. In this section, we briefly discuss other scenarios, which may also result in an upturn behavior in $B_{c2}(T)$, such as the Klemm-Luther-Beasley (KLB) model, the two-band superconductivity, the Takahashi-Tachiki effect, and the Ising superconductivity.

### (1) Klemm-Luther-Beasley model

The KLB model describes the temperature dependent in-plane critical field of layered superconductors, where an upturn behavior may be observed due to the dimensional crossover[15-17]. We have attempted to fit our experimental data using the KLB formula, but yields an unphysically short spin scattering time of 4.3 fs, which is even shorter than the total scattering time of 18.8 fs.

Furthermore, according to the KLB model, the disorder in the layered superconductors can enhance the spin-orbit scattering and randomize the spins of the electrons, which protects the superconductivity under large in-plane magnetic field. Thus, the $B_{c2,\parallel}$ is expected to be larger in the samples with stronger disorder. However, for $Ba_6Ta_{11}S_{28}$ superlattice, the $B_{c2,\parallel}$ becomes smaller and the upturn behavior in $B_{c2,\parallel}(T)$ is largely suppressed with increasing the disorder level (Fig. 4b in the main text), which is contradict to the KLB model. Therefore, the KLB model cannot explain our observations.

### (2) Two-band superconductivity

An upturn behavior in $B_{c2,\parallel}(T)$ may also appear in disordered two-band superconductors due to the intraband and interband scattering[18]. Within the two-band scenario, the upturn behavior appears in both $B_{c2,\parallel}(T)$ and $B_{c2,\perp}(T)$. In contrast, the pronounced upturn behavior in $Ba_6Ta_{11}S_{28}$ is limited in a small angle range $\Delta\theta_c = |\theta - 90°| \leq 0.5°$, which cannot be described by the two-band model. Furthermore, the $B_{c2}$ enhancement in the disordered two-band superconductivity model has a negative correlation with the intraband diffusivities, which decrease with stronger disorder. Thus, with increasing disorder, $B_{c2}$ is expected to be enhanced in this model, rather than suppressed as observed in $Ba_6Ta_{11}S_{28}$. Therefore, the two-band superconductivity can be ruled out as a possible scenario for the upturn behavior of $B_{c2,\parallel}(T)$ in $Ba_6Ta_{11}S_{28}$.

### (3) Takahashi-Tachiki effect

The Takahashi-Tachiki effect is proposed to describe the in-plane critical field in a superlattice composed of two superconducting layers with similar $T_c$ but different diffusion constants[19,20]. This model can also be applied to a disordered superconductor/normal-metal superlattice. However, in our system, $Ba_6Ta_{11}S_{28}$ single crystal consists of stacked superconducting layers (H-$TaS_2$) and insulating layers ($Ba_3TaS_5$). The density of states (DOS) of $Ba_6Ta_{11}S_{28}$ around the Fermi level are contributed by the TaS$_2$ layer and the DOS in the Ba$_3$TaS$_5$ layer is close to zero, evidenced by the ARPES results and band structure calculations[7]. In the limit $N_N / N_S \to 0$ (where $N_S$ and $N_N$ are the DOS of the superconducting and normal layers, respectively), the Takashi-Tachiki model cannot lead to the upturn behavior and thus cannot explain our observations.

Moreover, based on the Takahashi-Tachiki effect, the $B_{c2}$ is expected to be larger with smaller diffusion constant, which decreases with stronger disorder. Thus, with increasing disorder, $B_{c2}$ is expected to be enhanced in this model, rather than suppressed as observed in Ba$_6$Ta$_{11}$S$_{28}$. Therefore, we can rule out the Takahashi-Tachiki effect as a possible origin for our observations.

**(4) Ising superconductivity**
Monolayer transition metal dichalcogenides (TMD) superconductors with the in-plane inversion symmetry breaking are identified as Ising superconductors, characterized by an enhanced in-plane critical field. Under the protection of strong Ising SOC, the in-plane critical field is suggested to exhibit an upturn behavior in the low temperature regime (the upturn temperature $T^* < 0.2T_c$)[21,22]. In Ba$_6$Ta$_{11}$S$_{28}$ superlattice, the pronounced upturn behavior is observed at higher temperature $T^* \approx 0.8T_c$. Thus, Ising superconductivity alone cannot explain our observations.

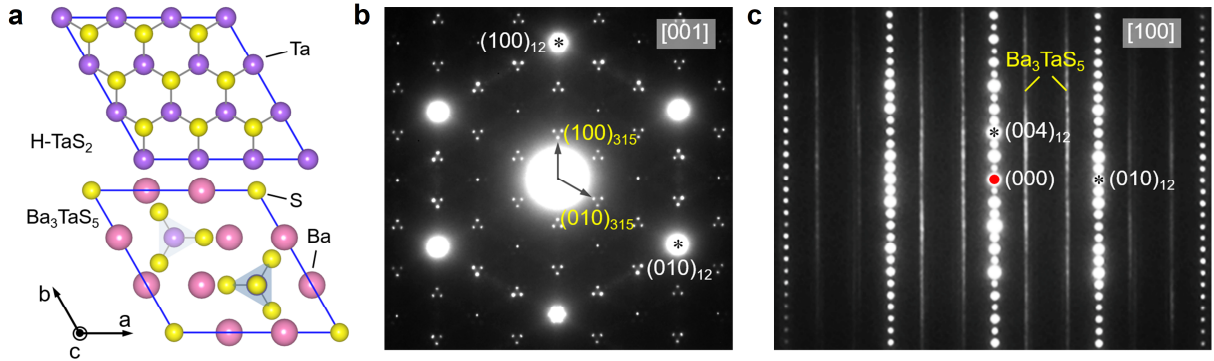

**Fig. S1 | Schematic and SAED patterns for $Ba_6Ta_{11}S_{28}$ single crystal. a**, Schematic illustration of the H-TaS$_2$ layer (top) and the Ba$_3$TaS$_5$ layer (bottom) along the [001] zone axis. **b**, SAED pattern taken along the [001] zone axis showing diffraction spots from H-TaS$_2$ layers (labeled by corner mark 12) and Ba$_3$TaS$_5$ layers (labeled by 315). The observed splitting of diffraction spots is attributed to the incommensurate lattice modulation. **c**, SAED pattern taken along the [100] zone axis of $Ba_6Ta_{11}S_{28}$ single crystal. The H-TaS$_2$ layers show sharp diffraction spots, labeled by corner mark 12, signifying a well-ordered lattice structure for the superconducting layers. The diffraction spots of Ba$_3$TaS$_5$ layers are observed to blur into elongated streaks, indicative of the structural disorder within the block layers.

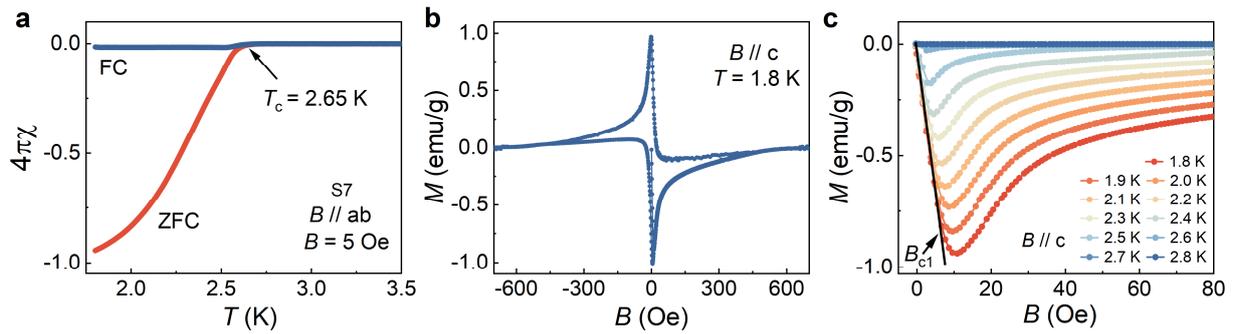

**Fig. S2 | Diamagnetic measurements on $Ba_6Ta_{11}S_{28}$ (sample S7). a**, Zero-field cooling (ZFC) and field cooling (FC) magnetic susceptibility $4\pi\chi$ under in-plane magnetic field of 5 Oe. **b**, Magnetization hysteresis loop at 1.8 K under out-of-plane magnetic field. The response is largely reversible, indicating weak vortex pinning. **c**, Magnetization $M$ as a function of field at different temperatures shown in the low field regime. $Ba_6Ta_{11}S_{28}$ crystal shows a clear diamagnetism below $B_{c1}$ due to the Meissner effect.

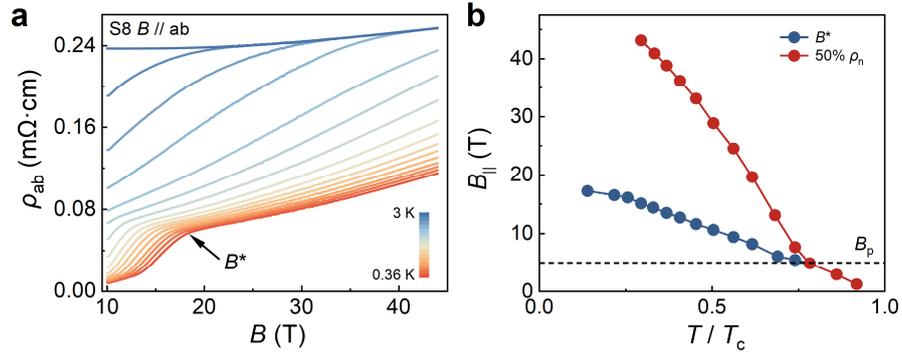

**Fig. S3 | The superconducting transition under steady high in-plane magnetic fields (sample S8). a**, The resistivity versus in-plane field measured in a hybrid magnet at different temperatures. The magnetic field range is 10-44 T for the hybrid magnet, with a superconducting magnet of 10 T and a water-cooled magnet of 0-34 T. **b**, Phase diagram of $Ba_6Ta_{11}S_{28}$ superlattice. A prominent upturn and the large enhancement of in-plane critical field are observed.

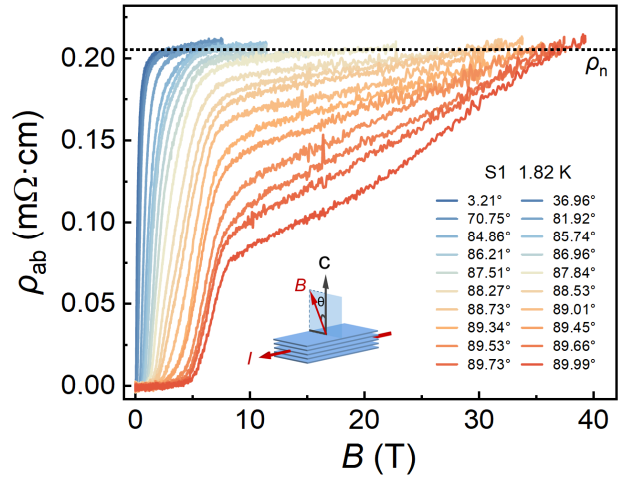

**Fig. S4 | The resistivity $\rho_{ab}$ versus magnetic field for different angles at 1.82 K (sample S1).** The measurements are conducted in a pulsed magnet.

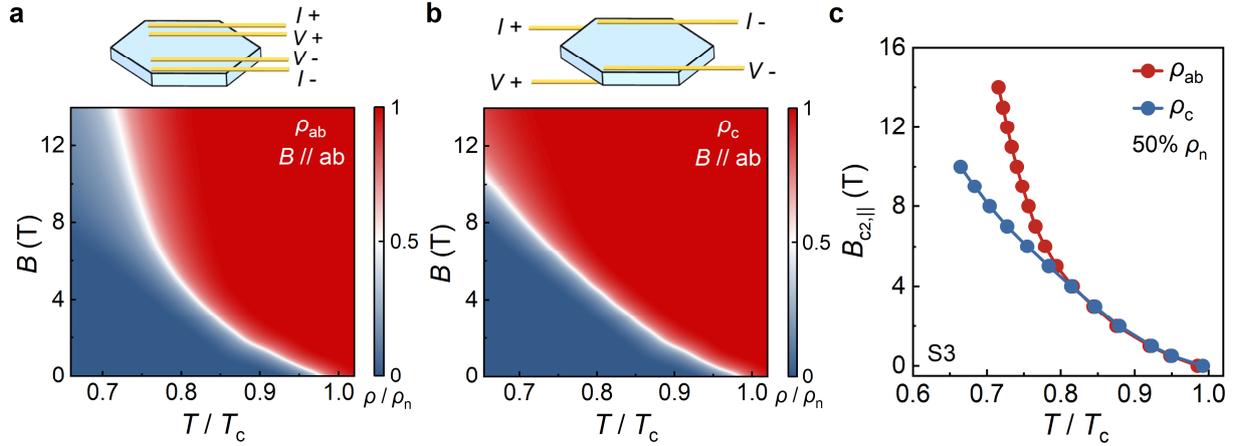

**Fig. S5 | Anisotropy between the resistivity along the intralayer ($\rho_{ab}$) and interlayer ($\rho_c$) directions under high in-plane fields (sample S3). a,b,** The color rendering of normalized intralayer (**a**) and interlayer resistivity (**b**) plotted as functions of reduced temperature and in-plane field. **c,** Comparison of the in-plane critical fields measured along the intralayer and interlayer directions. The temperature dependent in-plane critical field $B_{c2,\parallel}(T)$ for $\rho_{ab}$ shows a more pronounced upturn behavior compared to that for $\rho_c$.

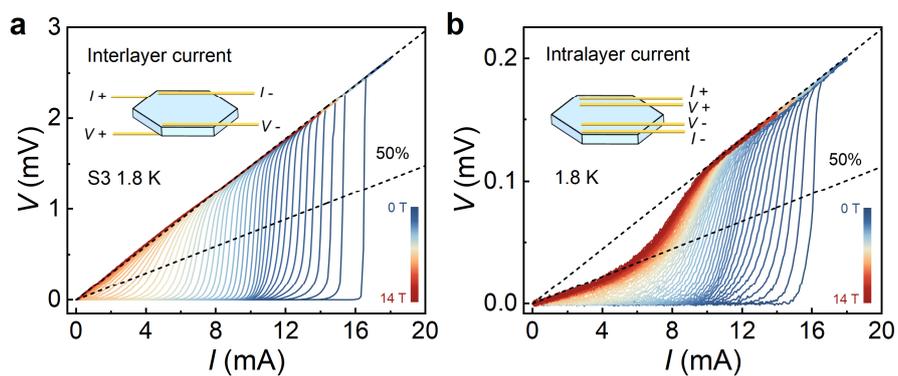

**Fig. S6 | *I-V* characteristics measured along the interlayer (a) and intralayer (b) directions at various in-plane fields at 1.8 K (sample S3).**

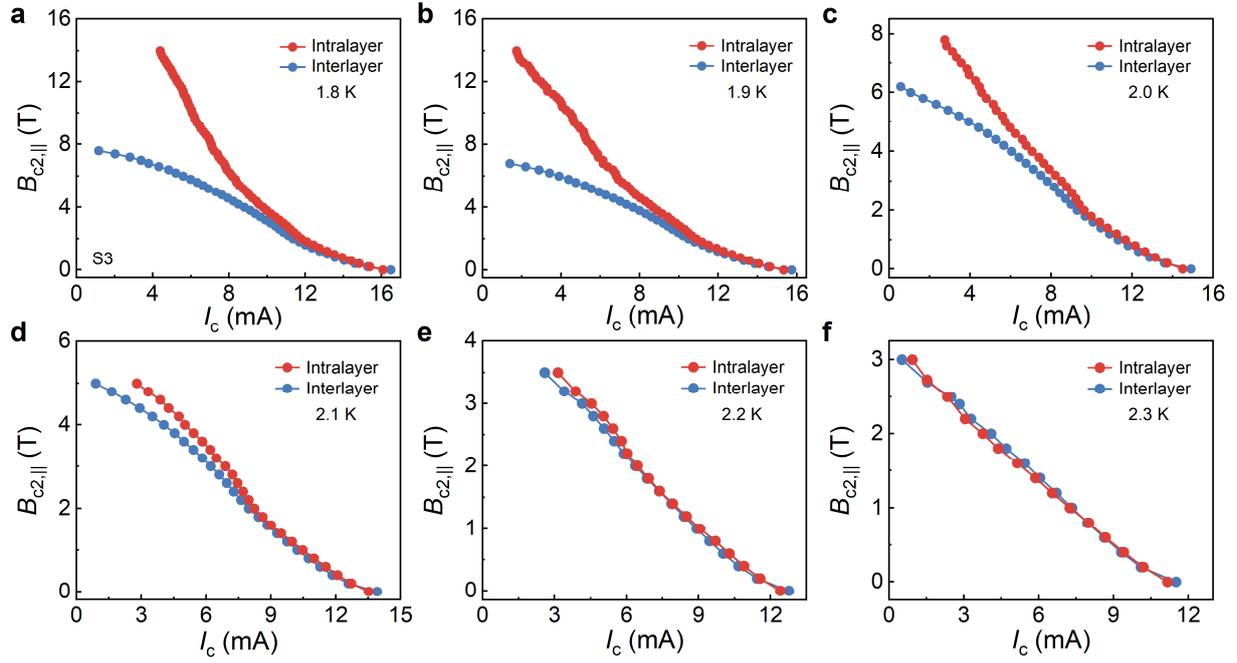

**Fig. S7 | The intralayer and interlayer critical currents under varying in-plane magnetic fields for sample S3 ranging from 1.8 K to 2.3 K.** The difference between the intralayer and interlayer critical currents gradually suppresses with rising temperature and disappears when $T \geq 2.3$ K.

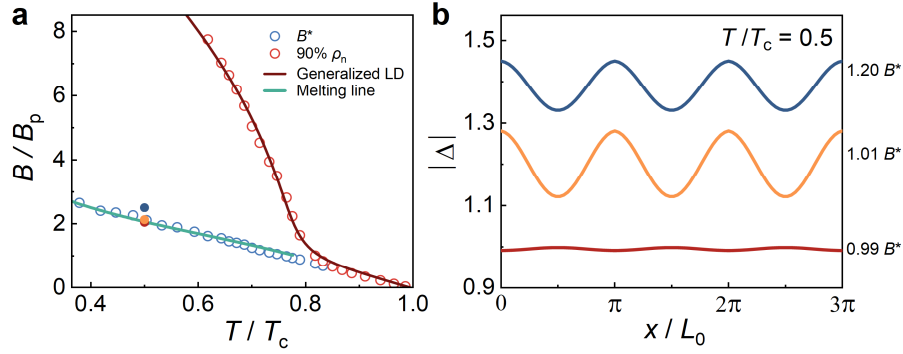

**Fig. S8 | Calculated orbital effect induced finite-momentum pairing state across the melting line of the Josephson vortex solid. a**, Fitting of the upper critical field and the melting line. The experimental data of the upper critical field and the characteristic field $B^*$ for Josephson vortex solid melting are represented by red and light blue circles, while the corresponding numerical fitting results are shown as dark red and cyan solid lines. The fitting of the melting line yields in-plane penetration length $\lambda_{ab}(0) = 1.2$ μm and Labusch parameter $\alpha_L = 2.5 \times 10^{-2}$ T$^2$/nm$^2$. **b**, Calculated spatial variation of the order parameter amplitude ($|\Delta|$) at $T/T_c = 0.5$ and $B/B^* = 0.99, 1.01$ and $1.2$ (corresponding to solid crimson, orange and blue circles in **a**), respectively. Here $B^*$ denotes the magnetic field where the melting occurs. For clarity, the curves are shifted vertically with the step of 0.2.

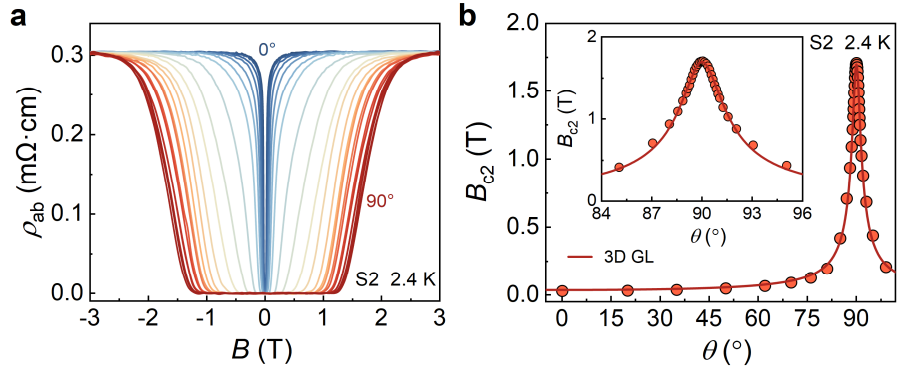

**Fig. S9 | Angular dependence of the critical field at 2.4 K (sample S2). a**, The resistivity as a function of magnetic field for different angles at 2.4 K. **b**, Angular dependence of the critical field $B_{c2}(\theta)$ at 2.4 K. The inset shows an enlarged view near 90°. $B_{c2}(\theta)$ can be fitted by the 3D Ginzburg-Landau formula for the entire angle range. Note that 2.4 K is higher than the temperature (2.3 K) where the $B_{c2,\parallel}(T)$ starts to show upturn behavior (see Fig. 4a in the main text for the phase diagram). The 3D superconducting nature above 2.3 K is also evidenced by the temperature dependence of the in-plane upper critical field in the high temperature regime.

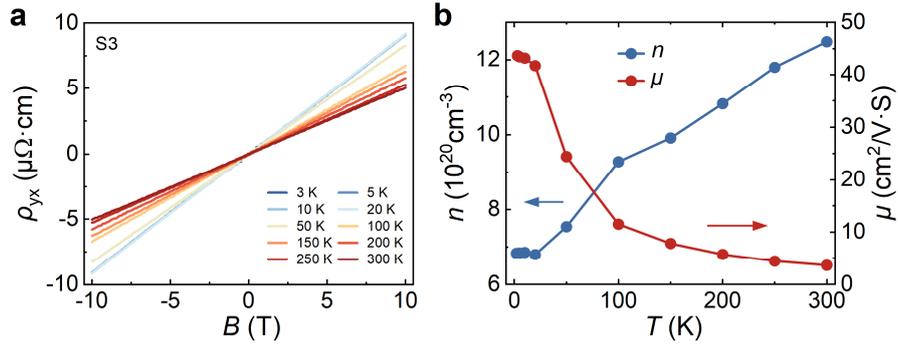

**Fig. S10 | Hall measurements on Ba$_6$Ta$_{11}$S$_{28}$ (sample S3). a**, Hall resistivity $\rho_{yx}$ as a function of field $B$ at various temperatures. **b**, The temperature-dependent carrier concentration $n$ and mobility $\mu$.

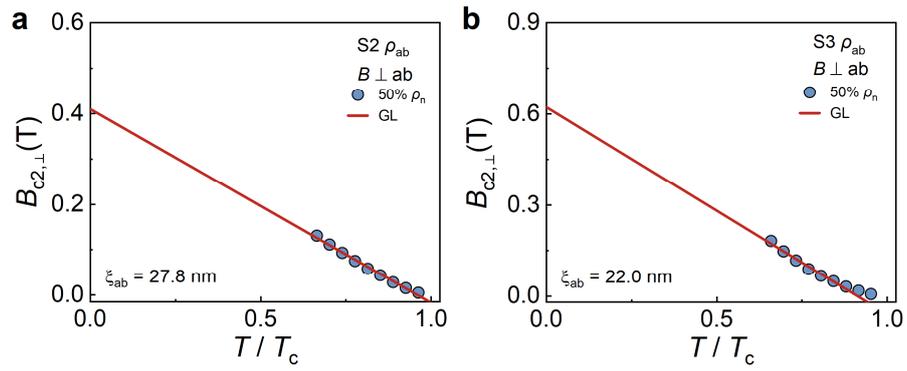

**Fig. S11 | Temperature dependence of the out-of-plane critical field in sample S2 (a) and sample S3 (b).** The red solid lines are the fitting curves of the GL formula.

**Table S1** | $Ba_6Ta_{11}S_{28}$ samples with different disorder levels. Higher disorder level is characterized by larger normal-state resistivity $\rho_{ab}(3\ \text{K})$ and smaller RRR.

| Samples | $\rho_{ab}(3\ \text{K})$ (mΩ·cm) | RRR | $B_{c2,\parallel}(2\ \text{K})$ (T) |
|---|---|---|---|
| S3 | 0.18 | 6.8 | 11.07 |
| S2 | 0.30 | 5.8 | 8.83 |
| S4 | 0.41 | 5.0 | 7.76 |
| S5 | 0.52 | 5.4 | 6.57 |
| S6 | 0.65 | 3.9 | 5.60 |

**Table S2** | The fitting parameters for $Ba_6Ta_{11}S_{28}$ samples with different disorder levels using the generalized Lawrence-Doniach model. The fitting parameter $B_0$ is the characteristic magnetic field scale corresponding to the Josephson vortex solid and $T_0$ is the characteristic temperature scale due to the interlayer Josephson coupling. The ratio of the effective thickness of the superconducting layer $d$ and the interlayer spacing $D$ is also presented in the table. The fitted ratio $d/D$ is found to be larger in the samples with higher disorder levels. With increasing disorder, the broadening of the superconducting state may originate from the less insulating $Ba_3TaS_5$ block layer.

| Samples | $B_0$ (T) | $T_0$ (K) | $d/D$ |
|---|---|---|---|
| S3 | 2.196 | 0.297 | 0.36 |
| S2 | 2.188 | 0.299 | 0.42 |
| S4 | 2.065 | 0.295 | 0.59 |
| S5 | 1.943 | 0.291 | 0.68 |
| S6 | 1.276 | 0.242 | 0.73 |